\documentclass[a4paper,11pt]{article}
\usepackage{pos}

\title{Bloch-Nordsieck restoration in $l \Bar{l} \rightarrow q \Bar{q}$}

\author*[a]{Franziska Reiner}
\author[a]{Axel Maas}

\affiliation[a]{Institute of Physics, NAWI Graz, University of Graz,\\
Universit\"atsplatz 5, A-8010 Graz, Austria}

\emailAdd{franziska.reiner@edu.uni-graz.at}
\emailAdd{axel.maas@uni-graz.at}

\abstract{Usually, the Bloch-Nordsieck theorem is expected to be violated in weak interactions, as the corresponding asymptotic states are not weak singlets. However, a manifest gauge-invariant approach replaces the asymptotic states with composite states, and thereby restores the Bloch-Nordsieck theorem. The Fr\"ohlich-Morchio-Strocchi mechanism ensures that this is in agreement with existing phenomenology. We outline the necessary steps of this construction.}

\FullConference{%
  *** The European Physical Society Conference on High Energy Physics (EPS-HEP2021), ***\\
  *** 26-30 July 2021 ***\\
  *** Online conference, jointly organized by Universität Hamburg and the research center DESY ***
}

\begin{document}
\maketitle

\section{Introduction}

In a perturbative approach to electroweak phenomenology, the asymptotic states are taken to be literally the elementary states of the Lagrangian. But because these states are not weak singlets the Bloch-Nordsieck theorem \cite{Bloch:1937pw} is violated, which leads to a substantial impact on cross sections at high energies \cite{Ciafaloni:2000rp,Bauer:2018xag}.

From a field-theoretical point of view the use of the elementary particles as asymptotic states is fundamentally flawed due to their gauge-dependence \cite{Frohlich:1981yi}. In fact, even in presence of a Brout-Englert-Higgs effect the asymptotic states need to be gauge-invariant, composite states \cite{Frohlich:1981yi}, which do not carry non-Abelian charges. This appears to be at first surprising, as the vast phenomenological success of ignoring this problem seems to indicate. This is resolved due to a combination of the Fr\"ohlich-Morchio-Strocchi (FMS) mechanism and the special structure of the standard model.

A general overview of this can be found in \cite{Maas:2017wzi}. Here, it will be outlined how this restores the Bloch-Nordsieck theorem for a lepton collider, and estimates the impact of this on the cross-section for  $l \Bar{l} \rightarrow q \Bar{q}$.

\section{The FMS mechanism}

In the following, we concentrate on the initial-state leptons. Up to strong-interactions effects, the procedure is the same for the final-state quarks. The standard model leptons\footnote{Hypercharge/electromagnetism plays a mute role for the following. If desired, details on how to include them can be found, e.\ g., in \cite{Maas:2017wzi}. For simplicity, we assume as well the existence of a right-handed neutrino.} form either a left-handed weak doublet $\psi_L=(\nu_L,l_L)^T$ or two right-handed singlets, $\nu_R$ and $l_R$.

The left-handed leptons are not gauge singlets. To build a gauge-invariant composite state with the same $J^{PC}$ quantum numbers we utilize the Higgs field $\phi$ before employing the Brout-Englert-Higgs effect and written as a matrix
\begin{equation}
    X=\left(\begin{array}{cc}
    \phi_{2}^{*} & \phi_{1} \\
    -\phi_{1}^{*} & \phi_{2}
    \end{array}\right).
\end{equation}
This field transforms under weak gauge transformations by a left-multiplication and under custodial transformation by a right-multiplication. Thus, the object \cite{Frohlich:1981yi}
\begin{equation}
    \Psi_{L}=X^{\dagger} \psi_{L}\label{fbs}
\end{equation}
is a gauge-invariant, spin 1/2 custodial doublet. Note that the right-handed fermions do not need to be dressed.

While it is always field-theoretically valid to write down this object, it appears at first rather similar to a hadron, and it is not at all obvious that it resembles anything present in the standard model. This is achieved by the FMS mechanism \cite{Frohlich:1981yi}. The poles of the propagator $\left\langle\overline{\Psi}_{\mathrm{L}}^i \Psi_{\mathrm{L}}^j\right\rangle$ will yield the mass of the object. To determine them, expand the Higgs field around its vacuum expectation value $v$ in a suitable gauge, $\phi=v+\eta$, where $\eta$ is the fluctuation field. This yields
\begin{eqnarray}
    \left\langle\overline{\Psi}_{\mathrm{L}}^1 \Psi_{\mathrm{L}}^1\right\rangle&=&v^{2}\left\langle\overline{\mathrm{\nu}_{\mathrm{L}}} \mathrm{\nu}_{\mathrm{L}}\right\rangle+\mathcal{O}(\eta/v)\\
    \left\langle\overline{\Psi}_{\mathrm{L}}^2 \Psi_{\mathrm{L}}^2\right\rangle&=&v^{2}\left\langle\overline{\mathrm{e}_{\mathrm{L}}} \mathrm{e}_{\mathrm{L}}\right\rangle+\mathcal{O}(\eta/v),
\end{eqnarray}
Thus, to this order the masses of both doublet members are the ones of the elementary electron and neutrino. This mechanism is general, and has been extensively tested on the lattice \cite{Maas:2017wzi}, especially also for fermions \cite{Afferrante:2020fhd}, and is supported non-perturbatively. Furthermore, this expansion can be systematically put into a perturbative framework \cite{Maas:2020kda}.

\section{The process $l \Bar{l} \rightarrow q \Bar{q}$}

By its very construction, it is clear that a scattering process only involving (\ref{fbs}) necessarily respects the Bloch-Nordsieck theorem. It is nevertheless very interesting how this works out, and especially why this process does not deviate at low, i.\ e.\ LEP2, energies appreciably from the ordinary result involving elementary particles as asymptotic states. We suppress here several formal aspects, which will be discussed elsewhere \cite{Reiner:unpublished}.

Since the violation of the Bloch-Nordsieck theorem has at leading-double-log order only implications for the scattering of two-left-handed particles \cite{Ciafaloni:2000rp}, we will concentrate on this case. The corresponding matrix element is then
\begin{equation}
\left\langle \bar{\Psi}_L^2\Psi_L^2 \Bar{Q}_L^2 Q_L^2\right\rangle=\left\langle\bar{\psi}_L^i X_{i2}X^\dagger_{2j}\psi_L^j\bar{q}_L^k X_{k2}X^\dagger_{2l}q_L^l\right\rangle\label{me}
\end{equation}
where $Q$ is the composite state for the left-handed top-type (and bottom-type) quark constructed analogously to (\ref{fbs}) from  $q=(b,t)^T$. We concentrate here on the would-be $\bar{l}l\to\bar{t}t$ process, and select the custodial indices accordingly to be 2. Calculating (\ref{me}) is a formidable task. However, we can analyze here two special cases.

First, we investigate what happens when performing the FMS mechanism in leading order in $v$. In this case, $X=v1$, and the expression collapses as
\begin{equation}
\left\langle \bar{\Psi}_L^2\Psi_L^2 \Bar{Q}_L^2 Q_L^2\right\rangle=v^4\left\langle \bar{l}_L l_L \bar{t}_L t_L\right\rangle+\mathcal{O}(\eta/v)
\end{equation}
Thus, to this order, we recover the usual expression for the matrix element. This shows how the FMS mechanism recovers the usual result if higher orders in $v$ can be neglected. Since the process is primarily characterized by the $s$-channel, the expected dimensionful quantity characterizing the validity is $s/v$ \cite{Egger:2017tkd}. Thus, at LEP2 energies this should be a fairly good approximation. However, the Bloch-Nordsieck violation also plays no role at these energies \cite{Ciafaloni:2000rp,Bauer:2018xag}, so it is no contradiction that the violation appears to be recovered.

This can therefore no longer be the case if $s/v\gg 1$, like at future linear colliders. Assuming a fully exclusive measurement, however, it is safe to assume that the Higgs fields in (\ref{me}) act primarily as spectators. This is also supported by investigations of the substructure of the bound state (\ref{fbs}) \cite{Egger:2017tkd,Afferrante:2020fhd}. This is akin to the situation in hadron scattering where the other partons are also spectators. In this case, like in a PDF construction, we thus have to sum the cross-sections of the possible pairings. Ignoring for a moment the final state, this implies that the total cross section is given by
\begin{equation}
\sigma_{\bar{\Psi}_L^2\Psi_L^2\to X}\sim\sigma_{\bar{l}_L l_L\to X}+\sigma_{\bar{l}_L \nu_L\to X}+\sigma_{\bar{\nu}_L l_L\to X}+\sigma_{\bar{\nu}_L \nu_L\to X}\label{tsx}.
\end{equation}
Now, each of the individual cross sections violates the Bloch-Nordsieck theorem, which leads to a double-logarithmic Sudakov enhancement \cite{Ciafaloni:2000rp}. If $s\gg v$, they become
\begin{eqnarray}
\sigma_{\bar{l}_L l_L\to X}&=\sigma_{\bar{\nu}_L \nu_L\to X}=&A+S\\
\sigma_{\bar{l}_L \nu_L\to X}&=\sigma_{\bar{\nu}_L l_L\to X}=&A-S,
\end{eqnarray}
where $A$ contains the non-enhanced part, and $S$ the enhancement proportional to the Sudakov logarithm squared. Inserting this into (\ref{tsx}) immediately shows that the enhancement cancels. The decisive step here is that it is necessary to do the summation, as mandated by gauge invariance. That they would cancel if summed follows from the Bloch-Nordsieck theorem, and is well known \cite{Ciafaloni:2000rp}. Thus, the FMS mechanism explains that the summation has to happen, but that this does not alter the low-energy behavior. Of course, this is now the behavior at small and large energies compared to the Higgs vacuum expectation value. For the intermediate range $s\approx v$ it will be necessary to study the expression (\ref{me}) which, while straightforward, is tedious.

\section{Summary and outlook}

The impact of the violation of the Bloch-Nordsieck theorem has been estimated to be of the same order as the leading-order strong correction at $s=1$ TeV, even in the unpolarized case  \cite{Ciafaloni:2000rp}. After the restoration of the Bloch-Nordsieck theorem, this contribution would no longer be present. This is a substantial modification of the cross-section. Depending on polarization, this effect can be enhanced or diminished, as once one participant is right-handed the effect is essentially eliminated \cite{Ciafaloni:2000rp}. This would suggest running a future lepton collider also in a left-left polarization mode to test this prediction. This would be particularly valuable as for many new physics scenarios even qualitative impacts would be expected \cite{Maas:2017wzi}, and thus experimentally confirming the field-theoretical foundation in such a way is crucial. On the other hand, it would substantially alter our understanding of the field theory of the standard model, if the predicted restoration of the Bloch-Nordsieck theorem would not take place. Moreover, similar effects are expected to play important roles also in many other processes than just the one we discussed here \cite{Bauer:2018xag}. This will be explored in more detail elsewhere \cite{Reiner:unpublished}.\footnote{We are grateful to Simon Pl\"atzer for a critical reading.}

\bibliographystyle{bibstyle}
\bibliography{bib}

\end{document}